# Quelles motivations à télétravailler ? Étude exploratoire en contexte post-confinement

**Steve BERBERAT**
**Damien ROSAT**
**Armand Brice KOUADIO**

**Haute École Spécialisée de Suisse occidentale (HES-SO), HEG Arc, Neuchâtel, Suisse**

steve.berberat@he-arc.ch

**RÉSUMÉ**

Après l'excitation qui a suivi les premières mesures de télétravail induites par la Covid-19, de nombreuses problématiques de management sont très vite apparues. En particulier se sont posées les questions relatives à l'efficacité du travail des employés et aux conditions d'une adoption réussie de ce nouveau mode de fonctionnement. L'étude se concentre sur les questions suivantes : quel est le nouvel attrait au télétravail et quelles en sont les causes potentielles ? Dans quelles mesures les employés se sentent-ils maintenant aptes à réaliser leur travail à distance et pour quelles raisons ? Afin de répondre à ces questions, nous avons mené pendant plusieurs semaines, auprès d'employés (N = 170) exerçant dans des domaines variés, une enquête sur l'expérience, les aptitudes et les motivations au télétravail. Il en résulte que l'adoption du télétravail est effective pour les répondants. Ceux qui en ont fait l'expérience présentent un attrait très marqué ainsi qu'une aptitude à le pratiquer bien réelle, avec des freins technologiques presque inexistants. Le modèle théorique utilisé, qui s'appuie sur les modèles d'acceptation de la technologie (TAM), a permis de mettre en évidence un nouveau facteur important de l'adoption du télétravail : le gain de temps sur les trajets domicile-travail. La tendance qui se dessine vers une adoption durable du télétravail implique cependant un regard critique et une discussion sur son déploiement, à long terme, dans les organisations publiques et privées.

*Mots-clés* : télétravail, motivation, adoption, aptitude, Covid-19.

**ABSTRACT**

After the considerable excitement caused by COVID-19 and the first telework measures, many management issues became apparent and some questions quickly arose, especially about efficiency of employees' work and conditions for successful telework adoption. This study focuses on the following questions: what is the new interest to telework for employees and what are the potential reasons for this? How much employees feel able to do their work remotely now and why? To answer these questions, we conducted a survey over several weeks, that involved employees coming from different industries (N=170), in order to collect their experience, skills and motivations for teleworking. The results show that adoption of telework is real for the respondents. Those who have experienced it present a strong motivation and a real capacity to use it, with almost no technological barriers. The theoretical model that we used, based on Technology Acceptance Models (TAM), has highlighted an important new factor of telework adoption: time saved in commuting. The study points out that adoption of telework could be sustainable for both public and private organizations, and this requires a critical examination and discussion.

# 1    INTRODUCTION

Au lendemain de la pandémie de la Covid-19, le télétravail s'est très vite imposé comme le levier de flexibilité par excellence dans la plupart des organisations publiques et privées. Pour continuer à performer comme auparavant, ou alors atténuer les dégâts économiques liés l'impossibilité de mener une activité économique normale, les entreprises ont dû rivaliser d'inventivité afin de se maintenir à flot. En Suisse, respectant les recommandations du Conseil fédéral, certaines entreprises ont cessé tout ou partie de leurs activités alors que d'autres ont mis en place des solutions numériques permettant à leurs collaborateurs et collaboratrices de télétravailler.

En dehors de la joie de pouvoir continuer à travailler en restant chez soi, la question du management des performances est vite apparue en même temps que les nouvelles modalités de travail. Des modalités impliquant l'utilisation, et donc la prise en main d'une grande variété d'outils collaboratifs (Teams, Zoom, Khoo, WebEx, etc.). Comment assurer, dans de telles conditions, le contrôle à distance des résultats professionnels ? Quel modèle de coordination mettre en place ? Quel style de management instaurer ? Ou en ce qui concerne spécifiquement la présente communication, quelles motivations (pour les employés) à télétravailler ?

En effet, face à l'armada de nouveaux outils et débarrassés du regard inquisiteur du manager de ligne, les employés doivent trouver en eux-mêmes la motivation nécessaire pour continuer à remplir les attentes placées en eux. Cette performance individuelle représente en effet un jalon indispensable de la chaîne de valeurs menant vers les performances collectives et organisationnelles (Wood, 2018). Tout en responsabilisant l'individu, elle fait appel à des formes motivationnelles, notamment autorégulées, sur lesquelles l'entreprise a de moins en moins d'emprise (Ryan & Deci, 2006). S'il est aujourd'hui indéniable que les dimensions intrinsèque et extrinsèque restent indissociables de la motivation à travailler, une conception hybride l'emporte de plus en plus sur les approches manichéennes qui caractérisèrent les premiers travaux sur la motivation (Acatrinei, 2015; E. L. Deci & Ryan, 1985; Moon, 2000). Contraintes de maintenir leurs performances dans un environnement désormais fragmenté, les entreprises doivent désormais repenser les leviers motivationnels à mobiliser. Dans ce cadre, elles devront sans doute innover et imaginer de nouvelles façons de se coordonner et recomposer la "famille entreprise".

Or, une incertitude demeure quant à l'évolution de la pandémie et aux mesures drastiques éventuelles (par exemple l'apparition de nouvelles pandémies ou de crises similaires dans le futur), notamment un reconfinement. Indubitablement, ces situations sont synonymes de conséquences importantes et inconnues pour les entreprises. Elles soulèvent en outre la nécessité d'étudier les déterminants de l'adhésion au télétravail et les conditions de sa "réussite". Une réussite qui s'exprime ici en termes synergétiques dès lors que les dispositifs techniques déployés entrent en adéquation avec le cœur humain des organisations, en l'occurrence le bien-être des utilisateurs (Emery, Boukamel, & Kouadio, 2019; Emery, Giauque, & Gonin, 2019; Van De Voorde, Veld, & Van Veldhoven, 2016).

Cette étude porte un regard critique sur le télétravail. Après une année de déploiement, le recul est suffisant pour en explorer les premiers effets. Il s'agit ici d'analyser, d'une part, les facteurs de motivations et d'aptitudes au télétravail et, d'autre part, l'impact que le semi-confinement a eu sur cette pratique. Aussi, elle met en lumière et ouvre la discussion sur les conséquences futures, pour les organisations et les employés, que la nouvelle tendance au travail à distance pourrait induire.

## 2 REVUE DE LA LITTÉRATURE

### 2.1 De la motivation au travail à la motivation à télétravailler

Le *moteur de l'action*, telle est la définition la plus générique que l'on puisse donner à la motivation après plusieurs décennies de recherche. La passion du sujet aidant, les chercheurs en management se sont emparés de ce sujet pour en faire un champ de recherche assez prolixe en lien avec le monde du travail. Ainsi qu'il s'agisse des théories du contenu ou des théories de processus, deux grandes formes de motivations paraissent désormais incontournables : la motivation intrinsèque et la motivation extrinsèque (Acatrinei, 2014; Emery 2012; Moon, 2000). Si de nombreuses études ont démontré le caractère durable de la motivation intrinsèque, il n'en demeure pas moins que les leviers concrètement mobilisés pour susciter la motivation au travail dépendent principalement de la conception qu'on se fait de l'homme en situation de travail (Chanlat, 2016; McGregor, 1960; Sainsaulieu, 2000) : soit partisan du moindre effort, cherchant à fuir le travail et les responsabilités ; ou alors coopératif et impliqué, trouvant dans le travail une source d'épanouissement. Selon le profil motivationnel de l'individu, les pratiques de gestion des ressources humaines pourront s'axer autour du contrôle plus ou moins strict des performances ou également être fondées sur la responsabilisation et l'autonomisation des employés.

Quel que soit le point de vue adopté, toute relation de travail s'inscrit dans le double mécanisme de l'échange, notamment social, et de l'adéquation de l'individu avec ses conditions, modalités et environnements de travail. En filigrane, une condition d'équilibre « contribution vs rétribution » qui renforce le sentiment d'équité et de justice. À défaut, l'employé sera amené à ajuster sa contribution (dans le sens de réduire ses efforts au travail) ; ajuster la perception de sa contribution (dénigrer son propre travail) ; ajuster la perception de sa rétribution (accorder davantage de valeur à ce qui est reçu) ; ou demander un ajustement de sa rétribution (Adams, 1965; Emery, Giauque, et al., 2019).

À terme c'est de prévention des comportements opportunistes de l'employé qu'il s'agit. Or, d'un point de vue empirique, contrôler les employés, notamment par le biais des instruments de management des performances, rencontre très vite des limites matérielles (coûts, faisabilité, biais). Des limites censées être comblées par des mécanismes incitatifs dont l'évolution s'est faite en parallèle de nos connaissances du phénomène motivationnel (Acatrinei, 2015; Delfgaauw & Dur, 2007; Wicht, 2017).

Si elle s'invite très vite au cœur de l'activité de télétravail, la prévention de ces comportements opportunistes, notamment par le contrôle technocratique du manager, devient caduque dès lors que les employés adhèrent à ce modèle de travail aujourd'hui imposé et sont motivés à en utiliser les outils. Cependant, tout en responsabilisant l'individu, le télétravail fait appel à des formes motivationnelles, notamment autorégulées, sur lesquelles l'entreprise a de moins en moins d'emprise (Ryan & Deci, 2006). Dans la théorie de l'autodétermination cela s'appelle la régulation intégrée, une référence au sens et aux besoins de réalisations personnelles (tout au sommet de la pyramide de Maslow). Dans cette optique, l'individu serait en quête de la valorisation de soi, en l'occurrence par le biais de ses performances en situation de télétravail (E. Deci & Ryan, 2004; Maslow, 2003).

Si l'impact du télétravail sur la motivation fait débat depuis maintenant une dizaine d'années dans les milieux académiques, la motivation à télétravailler est quant à elle plus récente (Caillier, 2012). La brève littérature sur le sujet montre en effet que la motivation à télétravailler dépend de plusieurs facteurs dont la capacité à faire bon usage de la technologie mobilisée, le style de management des performances adopté, la nature des tâches à réaliser, l'adéquation de l'environnement de travail domestique. Sur la base des travaux publiés ces dernières années, quelques traits essentiels de la motivation à télétravailler sont synthétisés ci-après.

Premièrement., le transfert de la responsabilité du management des performances vient avec nombre d'obstacles émotionnels dont une anxiété susceptible d'entraver l'efficacité en situation de télétravail (Maruyama & Tietze, 2012). Ces problèmes émotionnels sont majoritairement liés à l'isolement, la déconnexion, la désaffiliation ; lesquels résultent, entre autres, de l'épuisement dû aux interactions sociales dans le cadre du télétravail, sans oublier les difficultés de conciliation entre la vie professionnelle et la vie familiale. Des impacts évidents en matière de bien-être invitent à la prudence dans l'usage que l'on peut faire du télétravail (Anderson, Kaplan, & Vega, 2015; Lee, 2021). Deuxièmement, et en application de la théorie de l'action raisonnée, la motivation à adhérer et à utiliser les outils du télétravail dépend du contrôle que l'individu peut avoir de ses propres actions. Or s'agissant de la maîtrise individuelle de l'acte de télétravailler, plusieurs auteurs ont démontré que la facilité à être formé à l'outil augmente la motivation intrinsèque d'utiliser l'outil technologique (Ajzen & Fishbein, 1977; Pupion et al., 2017; Visawanath Venkatesh & Speier, 2000). Ceci même si le transfert des connaissances, préalable à des organisations plus flexibles, peut parfois représenter un défi dans la relation entre pratiquants ou non du télétravail (Taskin & Bridoux, 2010). Troisièmement, en lien avec le modèle *Job Demands-Resources*, il a été constaté que l'effet négatif du télétravail s'exprimait par une ambiguïté accrue des rôles et une réduction du soutien du management, qui à terme peut négativement impacter l'engagement au travail (Coenen & Kok, 2014; Delanoeije, Verbruggen, & Germeys, 2019).

Finalement, l'attitude des managers face au télétravail et leur capacité à manager à distance, respectivement à changer de paradigme managérial, contribuent au succès du déploiement du télétravail (Lee, 2021; Nunes, 2005; Taskin & Edwards, 2007). Autre facteur contribuant au succès du télétravail : le rôle clé du soutien technologique, et par extension du soutien organisationnel (Beauregard, Basile, & Canonico, 2019; Nakrosiene, Buciuniene, & Gostautaite, 2019; van Breukelen, Makkenze, & Waterreus, 2014). Dans la même veine, le type de travail et la nature des tâches à exécuter peuvent favoriser l'adhésion et l'efficacité dans le télétravail. Il est par exemple plus aisé pour des travailleurs du savoir d'en maîtriser plus facilement les rouages (Boell, Cecez-Kecmanovic, & Campbell, 2016).

En dehors des compétences ou des conditions de travail à proprement parler, la maîtrise des outils technologiques du télétravail est une autre donnée paradoxalement occultée. Comment atteindre en effet les performances individuelles et collectives escomptées si les technologies liées au télétravail ne sont elles-mêmes pas adoptées et convenablement utilisées ? La théorie de l'action raisonnée, appliquée au télétravail, montre que la maîtrise de l'outil de télétravail est un critère de motivation chez l'individu (Ryan & Deci, 2006), et cette condition est nécessaire pour qu'il passe d'une simple intention d'utiliser le télétravail à son utilisation réelle. Dès lors se pose la question des raisons initiales qui vont créer cette intention d'utiliser le télétravail chez l'individu.

## 2.2 De l'adoption des technologies au travail à l'adoption du télétravail

Les causes de l'intention d'utiliser une technologie sont depuis plus d'une vingtaine d'années abordées par le modèle d'acceptation de la technologie – *technology acceptance model* (TAM) (Davis, 1989). Il démontre l'effet indispensable de l'utilité et de la facilité d'utilisation perçues d'une technologie sur l'intention d'en faire effectivement usage. Chemin faisant, l'individu, ici le travailleur, s'adapte (et *in fine* adopte) à un nouvel environnement (technique) de travail. C'est le lieu de rappeler le double mécanisme assimilation-accommodation qui sous-tend tout développement intellectuel et qui s'applique ici à la notion d'adoption des technologies. Le premier de ces mécanismes, l'assimilation, suppose que l'organisme conserve sa forme. Inversement, le second mécanisme implique que l'organisme se modifie pour pouvoir intégrer la nouveauté (Piaget, 1953). Assimilation et accommodation procèdent non seulement de facteurs cognitifs, mais aussi affectifs qui tentent d'équilibrer un intérêt pour soi (dimension affective de l'assimilation) et un intérêt pour l'objet (dimension affective de l'accommodation) (Piaget, 1953).

Alors que beaucoup de recherches menées ces dernières années se sont focalisées sur la compréhension des variables externes expliquant l'intention d'utiliser une technologie ou plus spécifiquement l'utilité et la facilité d'utilisation perçues (Viswanath Venkatesh & Davis, 2000; Viswanath Venkatesh, Morris, Davis, & Davis, 2003), et que le courant de recherche de l'acceptation des technologies a été considéré comme mature (Viswanath Venkatesh, Davis, & Morris, 2007), une ultime mise à jour du modèle TAM a été faite. Dénommé TAM3, ce nouveau modèle synthétise les variables susceptibles d'influencer l'utilité et la facilité d'utilisation perçues d'une technologie (Viswanath Venkatesh & Bala, 2008).

Le modèle TAM3 est particulièrement pertinent pour expliquer l'adoption des technologies et plus spécifiquement des outils de télétravail. D'une part, il démontre l'influence de 12 variables précises sur l'utilité et la facilité d'utilisation perçues d'une technologie. D'autre part, il a permis d'expliquer de manière efficace l'attitude des cadres intermédiaires face à l'adoption du télétravail en entreprise (Silva-C, 2019). Ceci montre que fondamentalement, le modèle TAM pourrait s'appliquer en dehors du cadre *stricto-sensu* de l'adoption des technologies informatiques. Spécifiquement, cette étude menée par Silva-C valide 19 hypothèses mettant en relation plusieurs variables explicatives de l'attitude à utiliser le télétravail. Les auteurs présentent ces résultats sous la forme d'un modèle structurel que nous avons reproduit, traduit et réagencé dans la Figure 1 ci-après. Les résultats montrent le rôle important de l'expérience du télétravail dans le développement d'une attitude à utiliser le télétravail, ceci en impactant plusieurs variables intermédiaires telles que la compatibilité perçue de son emploi avec le télétravail, ou encore la motivation intrinsèque à télétravailler.

Les résultats de Silva-C (2019) montrent donc que ce ne sont pas uniquement des facteurs technologiques qui influencent l'attitude à réellement télétravailler. Aguilera, Lethiais, Rallet & Proulhac (2016) avaient par ailleurs déjà mené une étude auprès de PME françaises qui mettait en évidence l'importance particulière des facteurs organisationnels, comparativement aux dispositifs technologiques. Cette étude, outre le fait d'avoir révélé le télétravail comme une pratique souvent tacitement (et non formellement) acceptée en France, mettait en avant 4 types de freins : (1) l'incompatibilité des pratiques de télétravail avec le travail des répondants ; (2) l'incompatibilité du télétravail avec les méthodes de management ; (3) le possible déficit de productivité ainsi que (4) la barrière technologique qui, bien que moins importante comparativement aux facteurs organisationnels, a tout de même été considérée comme « forte » dans cette étude (Aguilera, Lethiais, Rallet, & Proulhac, 2016).

Figure 1 – Reproduction, traduction et réagencement des résultats de Silva-C (2019)

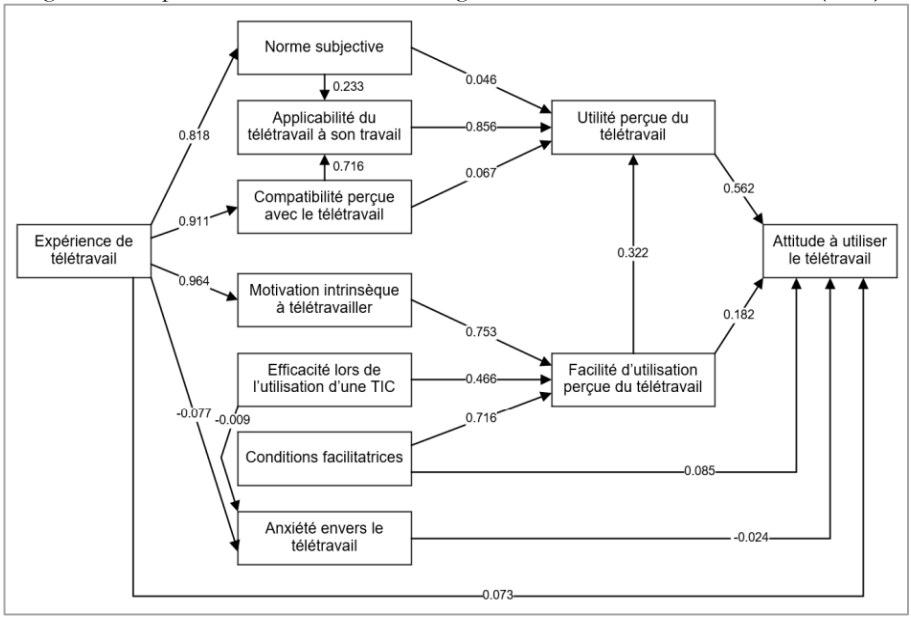

Malgré ces résultats intéressants, l'étude de Silva-C (2019) ne prend pas en compte l'ensemble des facteurs liés à d'adoption du télétravail, en l'occurrence les facteurs organisationnels. À ce titre, des rapports récents d'enquêtes menées auprès des cadres de l'administration publique française pointent l'insécurité de ces derniers : d'abord en matière de soutien organisationnel et social ; ensuite dans la coordination du travail à proprement parler, notamment la perte de l'aide des collègues, le fait de se voir moins impliqués dans les projets et autres actions collectives organisées par les représentants du personnel ; enfin le risque plus prononcé de dépression chez cette catégorie d'employés comparativement aux non-télétravailleurs (Hallépée & Mauroux, 2019). En plus de ces facteurs organisationnels, l'étude de Silva-C (2019) se concentre sur les facteurs jugés pertinents chez les cadres intermédiaires, ce qui écarte d'emblée ceux spécifiques aux employés subalternes.

L'arrivée de la Covid-19 avec son lot de périodes de confinement ou semi-confinement a fortement incité, voire contraint les entreprises à pratiquer du télétravail. Il nous paraît important d'en comprendre les impacts et les bouleversements induits. Dans cette optique, la présente recherche engage la discussion et apporte des réponses aux questions suivantes : quel est le nouvel attrait au télétravail et quelles en sont les causes potentielles ? Dans quelles mesures les employés se sentent-ils maintenant aptes à réaliser leur travail à distance et pour quelles raisons ? En outre, l'étude veut aussi apporter un éclairage sur d'éventuels nouveaux éléments à prendre en compte dans les modèles d'adoption du télétravail qui s'appuient sur le TAM.

## 3   CADRE THÉORIQUE & HYPOTHÈSES DE RECHERCHE

Le modèle d'adoption du télétravail proposé par Silva-C (2019), reproduit en Figure 1, sert de référence pour analyser le nouvel attrait du télétravail et le sentiment d'aptitude à télétravailler chez les employés. Nous en reprenons les trois variables principales héritées de longue date des modèles d'acceptation de la technologie (Viswanath Venkatesh & Bala, 2008; Viswanath Venkatesh et al., 2003; Visawanath Venkatesh & Speier, 2000) : l'utilité perçue du télétravail ; sa facilité d'utilisation perçue et l'attitude à l'utiliser, c'est-à-dire l'attitude démontrée par l'individu quant à son intention de pratiquer le télétravail. En conséquence, nous pouvons formuler les hypothèses suivantes :

- H1 : en contexte post-confinement, l'attitude à utiliser le télétravail dépend de l'utilité perçue du télétravail.
- H2 : en contexte post-confinement, l'attitude à utiliser le télétravail dépend de la facilité perçue du télétravail.
- H3 : en contexte post-confinement, l'utilité perçue du télétravail dépend de la facilité perçue du télétravail.

En plus de ces trois variables, nous ajoutons celle de l'expérience de télétravail. Elle nous semble particulièrement intéressante, car d'une part, elle est la seule indépendante dans le modèle de Silva-C (2019) et, d'autre part, une grande partie de la population s'est vue contrainte de faire l'expérience du télétravail. En nous inspirant des résultats obtenus par ces auteurs, nous formulons les trois hypothèses suivantes :

- H4 : en contexte post-confinement, l'utilité perçue du télétravail s'explique en partie par l'expérience préalable de télétravail.

- H5 : en contexte post-confinement, la facilité d'utilisation perçue du télétravail s'explique en partie par l'expérience préalable de télétravail.
- H6 : en contexte post-confinement, l'attitude à utiliser le télétravail s'explique en partie par l'expérience préalable de télétravail.

En complément de l'étude de Silva-C (2019), qui ne s'est concentrée que sur les cadres intermédiaires, nous proposons d'ajouter à notre cadre théorique, deux ensembles de facteurs que nous présumons volontairement comme étant inconnus à ce stade : premièrement les facteurs potentiels de motivation intrinsèque à télétravailler. Deuxièmement les facteurs potentiels d'aptitude à télétravailler. Ceci avec l'hypothèse que l'utilité perçue, ainsi que la facilité d'utilisation perçue dépendent de facteurs faisant partie de ces deux ensembles ; d'où les hypothèses suivantes :

- H7 : en contexte post-confinement, l'utilité perçue du télétravail dépend de facteurs de motivation intrinsèque à télétravail.
- H8 : en contexte post-confinement, la facilité d'utilisation perçue du télétravail dépend de facteurs de motivation intrinsèque à télétravail.
- H9 : en contexte post-confinement, l'utilité perçue du télétravail dépend de facteurs d'aptitude à télétravailler.
- H10 : en contexte post-confinement, la facilité d'utilisation perçue du télétravail dépend de facteurs d'aptitude à télétravailler.

La Figure 2 synthétise notre cadre conceptuel.

Figure 2 – Cadre conceptuel utilisé pour mener la recherche

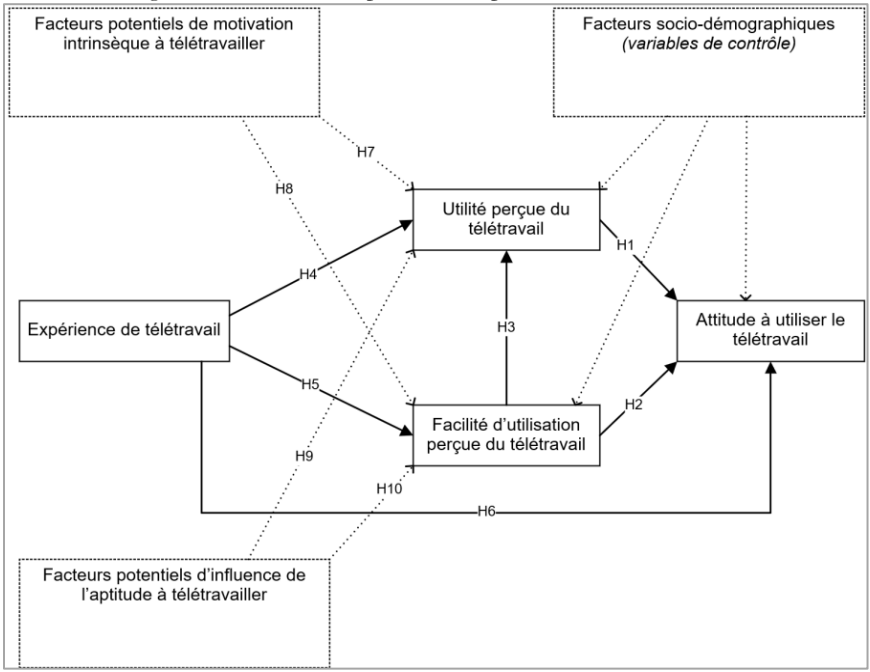

## 4 MÉTHODOLOGIE

La première phase de l'étude a consisté à faire une revue d'articles d'actualité grand public dans le but de recenser des facteurs pouvant potentiellement représenter la motivation intrinsèque à télétravailler ainsi que l'aptitude à télétravailler. Pour cela, la méthode de l'analyse thématique a été réalisée (Boyatzis, 1998). Cela nous a conduit à identifier de nouveaux facteurs potentiellement explicatifs de la motivation à télétravailler et de préciser les hypothèses sous-jacentes à H7, H8, H9 et H10. La seconde et principale phase de notre démarche empirique s'inscrit dans un paradigme positiviste, en l'occurrence une approche quantitative mobilisant un questionnaire à destination d'employées suisses romand (Comte, 1907; Creswell, 2008). Cette récolte de données avait pour but de vérifier les différentes hypothèses sous-jacentes à l'étude. Sans avoir utilisé un design probabiliste dans notre stratégie d'échantillonnage, les 170 réponses exploitables que nous avons obtenus (essentiellement récoltées via les médias sociaux en plus de réseaux professionnels) offrent une diversité intéressante devant permettre de croiser des points de vue de différents profils démographiques autour de l'expérience du télétravail (Paillé & Muchielli, 2003; Yilmaz, 2013).

Le logiciel R[1] est utilisé pour le traitement des résultats des questionnaires. Deux stratégies statistiques sont essentiellement mobilisées. Dans un premier temps, la matrice des corrélations a permis de filtrer et d'écarter les facteurs affichant une faible corrélation avec les trois variables principales suivantes : l'utilité perçue, la facilité d'utilisation perçue, ainsi que l'attitude à utiliser le télétravail. Dans un second temps, un modèle de régression linéaire a permis de rechercher et d'identifier les facteurs significatifs expliquant potentiellement chacune de ces trois variables, paramétrées alors comme variable réponse.

La démarche suivante est suivie pour chaque régression (Becker et al., 2016) :

1. Ajustement du modèle en sélectionnant toutes les variables explicatives identifiées selon le cadre théorique décrit en Figure 2, et en retirant celles qui ont pu être écartées préalablement avec la matrice de corrélations ;
2. Si au moins une variable explicative n'est pas significative (valeur de probabilité critique du test F, abrégée p-valeur, inférieure à 5%), réitération de la régression en supprimant la variable la moins significative ;
3. Lorsque toutes les variables restantes sont significatives, vérification de l'utilité (p-valeur < 5%) et de la qualité ($R^2$) de la régression ;
4. Analyse des diagnostics graphiques pour vérifier les conditions de normalité et d'homoscédasticité.
5. Vérification de l'aspect généralisation de la régression, en testant une à une les variables sociodémographiques de contrôle suivantes, de sorte à s'assurer qu'elles ne soient pas significatives dans le modèle : l'âge, le genre, le canton de domicile, celui du travail, la taille de l'organisation, le taux d'activité, le nombre d'enfants de 0 à 6 ans, le nombre d'enfants de 7 à 14 ans et enfin le nombre d'adultes dans le même domicile. Pour chacune d'elles, le test est effectué en ajoutant la variable en tant que facteur au modèle et en vérifiant sa significativité (p-valeur).

---

[1] R Core Team (2014). R: A language and environment for statistical computing. R Foundation for Statistical. Computing, Vienna, Austria. URL http://www.R-project.org

## 5  RÉSULTATS

### 5.1  Profil des répondants

Le profil des répondants correspond à une personne de 15 à 54 ans, ayant entre 0 et 3 enfants d'âges situés entre 0 et 14 ans, vivant parfois seule (27%) ou avec au moins un autre adulte dans le même domicile (73%). Cette personne peut être de genre masculin (62%) ou féminin (38%) et dispose d'une formation secondaire II ou tertiaire[2]. Elle est active professionnellement en Suisse romande et travaille autant dans des PME (57%) que dans de grandes organisations (43%).

### 5.2  Utilité, facilité d'utilisation et attitude à télétravailler

Après la période de semi-confinement du printemps 2020, une proportion de 73% des personnes interrogées percevait le télétravail comme étant utile. Alors que 18% se disaient neutre sur cet avis, 9% ne le considéraient pas utile.

Tout comme l'utilité perçue, la facilité d'utilisation du télétravail s'est révélée être bien présente chez les répondants. En effet, 84% d'entre eux qui considèrent que réaliser son travail à distance est ou serait facile, alors que seuls environ 10% sont en désaccord avec cet avis. Sur les aspects technologiques, ce sont 84% des personnes interrogées qui se disent tout à fait à « à l'aise » avec les technologies de télétravail contre seulement 0.6% qui sont en désaccord sur ce point. Enfin, pour 90% d'entre elles, les outils de communication étaient disponibles et en suffisance pour pouvoir télétravailler sur la fin du semi-confinement. Un manque vis-à-vis de ces outils a été souligné pour 3.8% des sondés.

L'attitude à utiliser le télétravail – reflétant la motivation globale à cette pratique – est bien présente chez les employés en période post-confinement. Ce sont véritablement 93% des répondants pour qui une évaluation positive de l'attitude à cette pratique a été mesurée. En outre, les réponses obtenues ont pu mettre en avant que 62% des interrogés souhaiteraient réaliser davantage de télétravail par rapport à ce qu'ils pratiquaient habituellement avant l'arrivée de la Covid-19. Pour les personnes désirant faire du télétravail, le taux idéal souhaité se situe majoritairement dans les plages 21-40% et 41-60% comme le montr

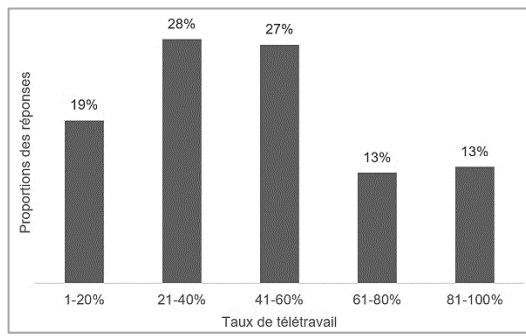

Figure 3 – Taux de télétravail idéalement souhaités

---

[2] https://systemeeducatif.educa.ch/fr [accès le 21.06.2021]

### 5.3 Causes expliquant l'attitude à télétravailler

Pour comprendre les causes pouvant expliquer l'utilité perçue, la facilité d'utilisation perçue ainsi que, finalement, l'attitude à télétravailler après le semi-confinement induit par l'arrivée du nouveau coronavirus, une matrice de corrélation a été effectuée sur l'ensemble des variables étudiées, suivi de différents tests de régression. Le Tableau 1 présente la matrice de corrélations obtenue. Les corrélations n'ont pas permis d'établir de potentiels liens cohérents entre certaines variables. Pour cette raison, aucune variable n'est écartée et chacune d'elles est incluse dans le processus de construction des régressions linéaires qui est opéré ensuite.

Tableau 1 – Matrice de corrélation de toutes les variables étudiées

|  | Âge | Aisance technologies télétr. | Attitude à utiliser le télétr. | Autonomie | Canton domicile | Canton travail | Compatibilité | Concentration | Conciliation vie privée-prof. | Expérience | Facilité utilisation perçue | Gain temps de trajets | Genre | Nb enfants de 0 à 6 ans | Nb enfants de 7 à 14 ans | Nécessité de réorg. travail | Réduction dérangements | Réduction stress et anxiété | Réduction volume réunions | Sentiment de disposer d'aide | Nombre d'adultes domicile | Suffisance outils de comm. | Taille organisation | Utilité perçue |
|---|---|---|---|---|---|---|---|---|---|---|---|---|---|---|---|---|---|---|---|---|---|---|---|---|
| Âge | 1.000 | 0.064 | -0.014 | 0.061 | 0.007 | -0.104 | 0.007 | 0.052 | -0.120 | 0.052 | -0.066 | 0.137 | 0.094 | 0.058 | 0.194 | -0.112 | 0.066 | -0.029 | 0.158 | -0.329 | -0.020 | 0.103 | 0.046 | -0.036 |
| Aisance technologies télétr. | 0.064 | 1.000 | 0.299 | 0.184 | -0.088 | -0.135 | 0.307 | 0.279 | 0.293 | 0.352 | 0.308 | 0.268 | -0.116 | 0.048 | 0.016 | -0.310 | 0.186 | 0.079 | 0.227 | -0.216 | -0.258 | 0.309 | 0.247 | 0.214 |
| Attitude à utiliser le télétr. | -0.014 | 0.299 | 1.000 | 0.380 | 0.027 | 0.088 | 0.367 | 0.408 | 0.345 | 0.398 | 0.469 | 0.364 | 0.032 | 0.095 | 0.029 | -0.330 | 0.012 | 0.248 | 0.252 | -0.253 | 0.041 | 0.305 | -0.046 | 0.463 |
| Autonomie | 0.061 | 0.184 | 0.380 | 1.000 | -0.010 | 0.172 | 0.140 | 0.225 | 0.164 | 0.323 | 0.373 | 0.045 | -0.031 | -0.300 | 0.033 | -0.048 | 0.161 | 0.600 | 0.342 | -0.008 | 0.003 | -0.013 | -0.093 | 0.194 |
| Canton domicile | 0.007 | -0.088 | 0.027 | -0.010 | 1.000 | 0.364 | -0.068 | -0.043 | -0.182 | 0.104 | -0.087 | -0.095 | 0.127 | 0.070 | 0.012 | 0.051 | 0.055 | -0.059 | 0.040 | 0.084 | -0.388 | -0.101 | -0.058 | 0.210 |
| Canton travail | -0.104 | -0.135 | 0.088 | 0.172 | 0.364 | 1.000 | -0.077 | 0.082 | -0.153 | 0.018 | -0.019 | -0.143 | 0.069 | -0.028 | -0.044 | 0.019 | 0.240 | 0.083 | 0.056 | -0.214 | -0.100 | -0.247 | 0.236 |
| Compatibilité | 0.007 | 0.307 | 0.367 | 0.140 | -0.068 | -0.077 | 1.000 | 0.181 | 0.108 | -0.015 | 0.070 | 0.174 | 0.148 | 0.052 | -0.008 | -0.288 | 0.059 | 0.043 | 0.181 | -0.227 | 0.158 | 0.602 | -0.061 | 0.137 |
| Concentration | 0.052 | 0.279 | 0.408 | 0.225 | -0.043 | 0.082 | 0.181 | 1.000 | 0.202 | 0.272 | 0.456 | 0.372 | -0.103 | 0.094 | -0.099 | 0.092 | 0.469 | 0.496 | 0.414 | -0.272 | -0.125 | 0.264 | 0.070 | 0.204 |
| Conciliation vie privée-prof. | -0.120 | 0.293 | 0.345 | 0.164 | -0.182 | -0.153 | 0.108 | 0.202 | 1.000 | 0.359 | 0.361 | 0.296 | -0.088 | 0.241 | 0.079 | -0.204 | 0.079 | 0.326 | 0.227 | -0.108 | 0.055 | 0.347 | 0.234 | 0.195 |
| Expérience | 0.052 | 0.352 | 0.398 | 0.323 | 0.104 | 0.018 | -0.015 | 0.272 | 0.359 | 1.000 | 0.421 | 0.100 | -0.164 | -0.014 | -0.052 | -0.033 | 0.103 | 0.321 | 0.345 | -0.019 | -0.206 | 0.069 | 0.049 | 0.479 |
| Facilité utilisation perçue | -0.066 | 0.308 | 0.469 | 0.373 | -0.087 | -0.019 | 0.070 | 0.456 | 0.361 | 0.421 | 1.000 | -0.047 | 0.215 | 0.137 | -0.208 | 0.115 | 0.473 | 0.311 | -0.390 | -0.104 | 0.219 | 0.115 | 0.263 |
| Gain temps de trajets | 0.137 | 0.268 | 0.364 | 0.045 | -0.095 | -0.143 | 0.174 | 0.372 | 0.296 | 0.100 | 0.234 | 1.000 | 0.091 | 0.080 | -0.021 | -0.131 | 0.044 | 0.113 | 0.056 | -0.074 | 0.026 | 0.265 | 0.077 | 0.191 |
| Genre | 0.094 | -0.116 | 0.032 | 0.031 | 0.127 | 0.069 | 0.148 | -0.103 | -0.088 | -0.164 | 0.031 | 0.234 | 1.000 | 0.199 | 0.133 | 0.123 | -0.191 | -0.179 | -0.044 | 0.153 | 0.003 | 0.003 | -0.015 | -0.091 |
| Nb enfants de 0 à 6 ans | 0.058 | 0.048 | 0.095 | -0.300 | 0.070 | -0.028 | 0.052 | 0.094 | 0.241 | -0.014 | 0.215 | 0.080 | 0.199 | 1.000 | 0.023 | -0.045 | -0.146 | -0.218 | -0.068 | -0.234 | 0.182 | 0.180 | 0.103 | 0.206 |
| Nb enfants de 7 à 14 ans | 0.194 | 0.016 | 0.029 | 0.033 | 0.012 | -0.044 | -0.008 | -0.099 | 0.079 | -0.052 | 0.131 | -0.021 | 0.133 | 0.023 | 1.000 | 0.179 | -0.359 | 0.074 | 0.066 | -0.134 | -0.031 | 0.263 | 0.185 |
| Nécessité de réorg. travail | -0.112 | -0.310 | -0.330 | -0.048 | 0.051 | 0.019 | -0.288 | 0.092 | -0.204 | -0.033 | -0.288 | -0.131 | 0.123 | -0.045 | 0.179 | 1.000 | -0.022 | -0.201 | 0.077 | 0.621 | -0.159 | -0.459 | -0.025 | -0.135 |
| Réduction dérangements | 0.066 | 0.186 | 0.012 | 0.161 | 0.055 | 0.240 | 0.059 | 0.469 | -0.079 | 0.103 | 0.115 | 0.044 | -0.191 | -0.146 | -0.359 | -0.022 | 1.000 | 0.224 | 0.486 | -0.100 | -0.177 | 0.020 | -0.055 | -0.090 |
| Réduction stress et anxiété | -0.029 | 0.079 | 0.248 | 0.600 | -0.059 | 0.083 | 0.043 | 0.496 | 0.326 | 0.321 | 0.473 | 0.113 | -0.179 | -0.201 | 0.074 | 0.224 | 1.000 | 0.270 | -0.044 | -0.061 | 0.164 | 0.058 | 0.162 | 0.185 |
| Réduction volume réunions | 0.158 | 0.227 | 0.252 | 0.342 | 0.040 | 0.056 | 0.181 | 0.414 | 0.227 | 0.345 | 0.311 | 0.056 | -0.044 | -0.068 | -0.043 | 0.077 | 0.486 | 0.270 | 1.000 | -0.053 | -0.224 | 0.057 | -0.026 | 0.244 |
| Sentiment de disposer d'aide | -0.329 | -0.216 | -0.253 | -0.008 | 0.084 | 0.056 | -0.227 | -0.272 | -0.108 | -0.019 | -0.390 | -0.074 | 0.153 | -0.234 | 0.066 | 0.621 | -0.100 | -0.061 | -0.053 | 1.000 | -0.164 | -0.053 | 1.000 | -0.405 | -0.061 | -0.095 |
| Nombre d'adultes domicile | -0.020 | -0.258 | 0.041 | 0.003 | -0.388 | -0.214 | 0.158 | -0.125 | 0.055 | -0.206 | -0.104 | 0.036 | 0.003 | -0.082 | -0.134 | -0.159 | -0.177 | 0.058 | -0.224 | -0.103 | 1.000 | 0.109 | -0.039 | -0.082 |
| Suffisance outils de comm. | 0.103 | 0.309 | 0.305 | -0.013 | -0.101 | -0.100 | 0.602 | 0.264 | 0.347 | 0.069 | 0.219 | 0.265 | 0.003 | 0.180 | -0.031 | -0.459 | 0.020 | 0.052 | 0.055 | 0.116 | -0.026 | -0.061 | 0.030 | 0.109 | 1.000 | 0.030 | -0.072 |
| Taille organisation | 0.046 | 0.247 | -0.046 | -0.093 | -0.058 | -0.247 | -0.061 | 0.070 | 0.234 | 0.049 | 0.161 | 0.077 | -0.015 | 0.031 | 0.263 | -0.025 | -0.055 | 0.116 | -0.026 | -0.061 | 0.030 | 1.000 | -0.072 |
| Utilité perçue | -0.036 | 0.214 | 0.463 | 0.194 | 0.210 | 0.236 | 0.137 | 0.204 | 0.195 | 0.479 | 0.263 | 0.191 | -0.091 | 0.206 | -0.002 | -0.135 | -0.090 | 0.185 | 0.244 | -0.095 | -0.082 | 0.086 | -0.072 | 1.000 |

Les modèles de régression ont, pour leur part, permis de valider les hypothèses H1 et H2 et H4. Les hypothèses H3, H5 et H6 n'ont pas pu être validés. En effet, les facteurs explicatifs se sont montrés non significatifs.

Tableau 2 – Significativités des facteurs potentiels de motivation à télétravailler

| Facteurs potentiels de motivation intrinsèque à télétravailler | Significativité (p-valeur) pour prédire : | |
|---|---|---|
|  | Utilité perçue du télétravail | Facilité d'utilis. perçue du télétr. |
| Autonomie dans l'organisation de ses tâches | 0.522 | 0.081 |
| Meilleure conciliation vie privée-profession. | 0.824 | 0.417 |
| Gain de temps sur les trajets | **0.003** | 0.453 |
| Stress et anxiété au travail réduits | 0.710 | **0.000004** |
| Meilleure concentration en télétravail | 0.329 | 0.177 |
| Nombre de dérangements réduit en télétr. | 0.907 | 0.171 |
| Volume des réunions réduit en télétravail | 0.572 | 0.059 |

Note : seuil de significativité p-valeur fixé 0.05

Le Tableau 2 présente l'ensemble des facteurs potentiels de motivation intrinsèque au télétravail identifiés, ainsi que les valeurs de significativités (p-valeur) obtenues. Rappelons que ces facteurs sont considérés comme expliquant potentiellement l'utilité perçue ainsi que la facilité perçue d'utilisation du télétravail. Deux variables se sont montrées significatives : le gain de temps sur les trajets, qui influence l'utilité perçue, ainsi que la réduction du stress et de l'anxiété au travail, qui a un effet sur la facilité d'utilisation perçue du télétravail. L'hypothèse H7 peut donc être validée par la variable du gain de temps sur les trajets. Quant à l'hypothèse H8, nous la retenons sur la base de la relation entre notre variable dépendante et les variables stress et l'anxiété au travail.

Le Tableau 3 fait de même pour les facteurs d'aptitude à télétravailler. Les résultats obtenus mettent en avant deux facteurs explicatifs significatifs parmi cet ensemble. Le premier indique que la compatibilité du travail de l'employé avec le mode de travail à distance influence l'utilité perçue du télétravail. Le second spécifie que le sentiment de disposer d'aide en cas de besoin pour télétravailler a un effet sur la facilité d'utilisation perçue du télétravail. Les hypothèses H9 et H10 sont donc validées. L'aisance dans les technologies de télétravail – seul facteur relatif aux compétences dans les technologies – ne s'est pas montré suffisamment significatif, avec une p-valeur de 6.2%.

Tableau 3 – Significativités des facteurs potentiels de l'aptitude à télétravailler

| Facteurs potentiels d'influence de l'aptitude à télétravailler | Significativité (p-valeur) pour prédire : | |
|---|---|---|
| | Utilité perçue du télétravail | Facilité d'utilis. perçue du télétr. |
| Compatibilité de son travail avec le télétr. | **0.0007** | 0.877 |
| Degré de nécessité de réorganiser son travail | 0.267 | 0.076 |
| Suffisance dans les outils de communication | 0.458 | 0.761 |
| Suffisance dans l'espace de travail à dispo. | 0.936 | 0.273 |
| Aisance dans les technologies de télétravail | 0.516 | 0.062 |
| Sentiment de disposer d'aide | 0.316 | **0.00009** |

Note : seuil de significativité p-valeur fixé 0.05

Tableau 4 – Significativités des variables de contrôle

| Variables sociodémographiques | Significativité (p-valeur) pour prédire : | | |
|---|---|---|---|
| | Attitude à utiliser le télétr. | Utilité perçue du télétr. | Facilité d'utilis. perçue du télétr. |
| Âge | 0.139 | 0.743 | 0.131 |
| Genre | 0.075 | 0.779 | 0.475 |
| Canton de domicile | 0.054 | 0.604 | 0.122 |
| Canton de travail | 0.307 | 0.364 | 0.530 |
| Taille de l'organisation | 0.076 | **0.046** | 0.330 |
| Taux d'activité | 0.099 | 0.212 | 0.223 |
| Nombre d'enfants de 0 à 6 ans | 0.993 | 0.174 | **0.032** |
| Nombre d'enfants de 7 à 14 ans | 0.963 | 0.413 | 0.468 |
| Nombre d'adultes dans le domicile | 0.557 | 0.442 | 0.172 |

Note : seuil de significativité p-valeur fixé 0.05

Les régressions linéaires ont été testées avec les différentes variables sociodémographiques de contrôles présentées dans le Tableau 4. Ces tests ont permis de révéler inopinément deux variables ayant une influence significative. En effet, il s'avère que la taille de l'entreprise fait partie des variables explicatives de l'utilité perçue, et que le nombre d'enfants âgés entre 0 et 6 ans détermine en partie la facilité d'utilisation perçue du télétravail.

L'ensemble des influences significatives de facteurs sur l'utilité perçue, la facilité d'utilisation perçue et l'attitude à utiliser le télétravail sont présentées sur le modèle de la Figure 4. Les valeurs sur les flèches représentent les coefficients des variables explicatives dans les régressions, et les étoiles (*) représentent leur niveau de significativité (p-valeur). Les coefficients sont donnés à titre indicatif et peuvent être comparés entre eux, puisque toutes les données utilisées dans les régressions ont été normalisées sur une même échelle de valeurs, allant de -2 à 2, correspondants respectivement à « pas du tout d'accord » et « tout à fait d'accord » de l'échelle de Likert. Enfin, les valeurs en pourcentage données sur les variables de l'utilité perçue, de la facilité d'utilisation perçue et de l'attitude à télétravailler représentent les valeurs $R^2$ ajustées obtenues avec les régressions.

Figure 4 – Modèle des facteurs de l'attitude à télétravailler

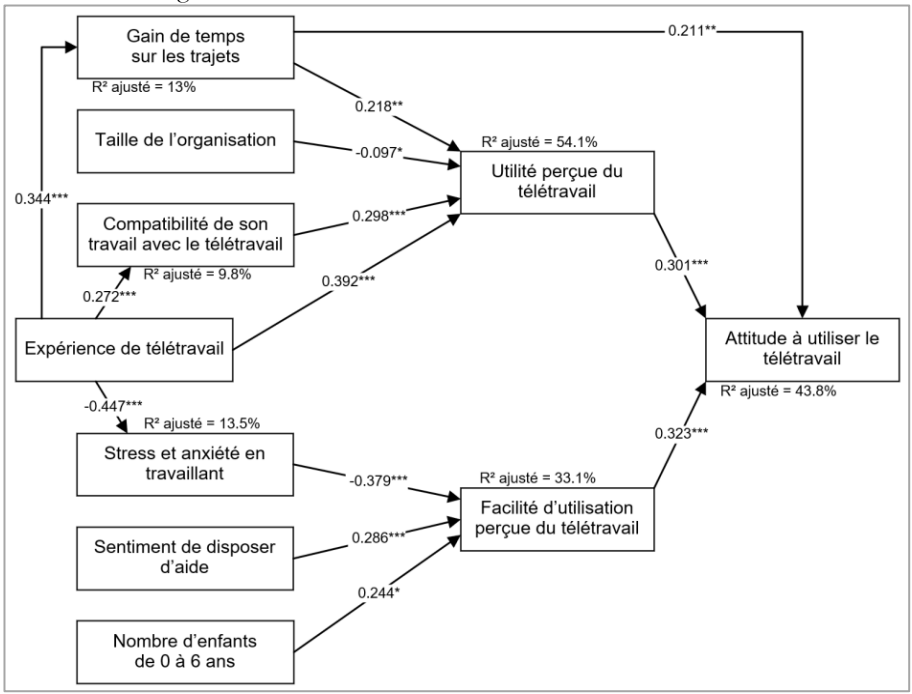

# 6 DISCUSSION
## 6.1 Interprétation des résultats

À la question de savoir « quel est le nouvel attrait au télétravail », les résultats révèlent, pour les employés interrogés, un nouvel attrait très marqué. L'indicateur de 93% de personnes motivées à cette pratique après l'expérience du semi-confinement souligne l'ampleur de ce phénomène. En plus d'un voyant qui semble être au vert concernant l'attrait au télétravail, celui de l'aptitude des employés à sa pratique se montre aussi favorable. Le chiffre de 84% de répondants, qui considèrent le télétravail comme une pratique facile d'utilisation et qui disent se sentir à l'aise dans l'utilisation des technologies y-relatives, apporte une réponse à la question de savoir « dans quelles mesures les employés se sentent-ils maintenant aptes à réaliser leur travail à distance ». Dès lors et en raison de l'expérience de télétravail qui a été faite, tout nouveau confinement qui pourrait avoir lieu dans le futur n'empêcherait pas la majeure partie des personnes interrogées sur le sujet de poursuivre leur activité professionnelle.

La littérature montre que l'attitude à adopter le télétravail dépend non seulement de sa facilité d'utilisation perçue, mais également de l'utilité que les individus y voient (Beauregard et al., 2019; Taskin & Edwards, 2007; Viswanath Venkatesh & Davis, 2000). Ici encore, le voyant est vert puisque 74% des répondants jugent le télétravail comme étant utile. Tout porte à croire que la pratique a un potentiel pour être généralisée ces prochaines années auprès des personnes suisses romandes dotées d'un niveau de formation secondaire II ou tertiaire.

Nous apportons également des réponses aux deux questions interrogeant les raisons potentielles de l'adoption du télétravail, à savoir « quelles sont les causes potentielles du nouvel attrait au télétravail ? » et « pour quelles raisons les employés se sentent-ils aptes au télétravail ? ». Les résultats révèlent que, pour les employés, l'attitude à télétravailler s'explique en partie par l'utilité et la facilité de le pratiquer qu'ils y perçoivent, ainsi que par un bénéfice direct pour eux qui est le gain de temps sur les trajets entre le lieu de travail et le domicile. La plupart des facteurs d'utilité et de facilité d'utilisation perçues confirment les résultats de travaux effectués par des instances officielles françaises (Hallépée & Mauroux, 2019), qui avaient entre autres déjà soulevé le besoin de disposer de soutien de la part de l'organisation et des collègues et qui apparaît dans nos résultats comme un facteur de facilité d'utilisation. Toutefois, contrairement aux résultats de ces études réalisées en France, il apparaît que les facteurs d'une plus grande autonomie dans la gestion des tâches ainsi que d'une meilleure conciliation entre vie privée et professionnelle ne sont pas déterminants de l'adoption du télétravail, que ce soit du point de vue de l'utilité ou de la facilité perçues.

Le facteur de gain de temps sur les trajets est en contradiction avec les résultats obtenus par l'étude menée par Aguilera et al. (2016), dans laquelle le temps de trajet domicile-travail y ressort comme non incitatif. Cela s'explique en partie par l'expérience forcée du télétravail induite par la Covid-19, qui semble avoir eu un effet révélateur de ce bénéfice chez certains employés. En effet, les résultats de notre étude montrent que l'expérience de télétravail est déterminante dans l'explication du gain de temps perçu sur les trajets. Le profil des populations cibles pourrait aussi s'ajouter à cette explication : ce gain de temps perçu pourrait, soit s'appliquer spécifiquement aux employés dotés d'un certain niveau de formation (en l'occurrence secondaire II ou tertiaire) ou soit, dépendre également d'une culture différente relative au temps de trajet entre les Français – public ciblé par l'étude d'Aguilera et al. (2016) – et les Suisses romands.

D'autre part, alors que les études de Aguilera et al. (2016) et de Silva-C (2019) confirment l'existence d'une barrière technologique significative, nos résultats montrent, au contraire, que l'efficacité de l'employé lors de l'utilisation des outils de télétravail – ou autrement dit sa capacité à en faire usage – n'est pas déterminante dans la facilité d'utilisation perçue et dans l'attitude à le pratiquer. Sommes-nous arrivés à un stade où la technologie peut appuyer un affranchissement des frontières physiques dans les activités professionnelles ? La réponse à cette question n'est que provisoire, toutefois les résultats apportés par la population visée par notre étude peuvent le laisser penser.

Concernant l'expérience de télétravail, les résultats montrent qu'elle joue un rôle pour déterminer la compatibilité du travail de l'employé avec le travail à distance, ainsi que la réduction du stress et de l'anxiété lorsqu'il est pratiqué. Ceci confirme les résultats obtenus par Silva-C (2019). Cette expérience ne s'est néanmoins pas révélée être un déterminant direct de l'attitude à télétravailler dans notre cas. Nous pensons que l'attitude à utiliser le télétravail pourrait *in fine* s'expliquer uniquement par des variables intermédiaires influencées par l'expérience de télétravail, et non de manière directe à l'image des résultats de Silva-C. Ces variables intermédiaires pourraient inclure, en plus de l'utilité et de la facilité d'utilisation perçues, les bénéfices personnels pour l'individu tels que le gain de temps sur les trajets. Par exemple le télétravail peut aussi apporter de nouvelles réponses à des problématiques d'aménagement du territoire dans les zones périurbaines. Mieux, des personnes en situation de handicap pourraient y trouver une opportunité pour améliorer les conditions de travail, voire s'insérer de façon plus aisée dans l'emploi[3] (Pouly, 2020).

Enfin, deux facteurs contre-intuitifs sont mis en lumière par les résultats. Le premier est celui de la taille de l'entreprise, qui semble avoir un effet négatif sur l'utilité perçue du télétravail. Le second est le nombre d'enfants entre 0 et 6 ans, qui influencerait positivement la facilité perçue à pratiquer le télétravail. À ce stade, nous ne disposons pas d'éléments susceptibles d'expliquer ces deux facteurs. Leur significativité n'étant pas très élevée (respectivement de 0.046 et 0.032), nous pensons que d'autres variables liées pourraient apporter plus d'explications. Une exploration de cet aspect au travers d'une étude qualitative serait nécessaire pour apporter des clarifications.

---

[3] https://www.legifrance.gouv.fr/jorf/id/JORFTEXT000035607366 [accès le 21.06.21]

Les résultats apportés par l'étude posent plusieurs questions, à commencer par celle de l'adoption à long terme du télétravail. DeLone et McLean (2003, 2016) soulèvent que, bien que l'adoption initiale d'un système d'information (SI) soit importante, une utilisation à long terme du SI est une mesure clé pour déterminer son succès. Leur modèle de succès – *information system success model* (ISSM) – montre que l'utilisation d'un SI et la satisfaction utilisateur ont un impact pour l'organisation, mesurable dans la plupart des cas sous forme de bénéfices nets tangibles ou intangibles, et que cet impact influence à son tour l'intention d'utiliser le système. Le succès des outils de télétravail pourrait donc également s'inscrire dans le modèle ISSM dès lors que l'on souhaite considérer leur utilisation à long terme.

Malgré les doutes subsistant quant à une adoption à long terme du télétravail, les résultats obtenus présentent néanmoins plusieurs signes en sa faveur : une utilité perçue par les employés, une facilité d'utilisation bien réelle avec des freins technologiques presque inexistants ainsi qu'une attitude et une motivation très forte à cette pratique. Un retour complet à la situation d'avant Covid-19 nous paraît peu réaliste. Cette adoption généralisée du télétravail permettrait aux organisations de revoir la conception de leurs locaux, en créant des espaces de *coworking* et en favorisant le partage des bureaux. Pour l'employeur, le bénéfice direct serait une réduction du volume des locaux et des charges y relatives. Des frais directs et indirects (loyer de l'habitat, eau, électricité, entretien, etc.) seraient *de facto* transférés de l'employeur vers l'employé. Enfin, la perte des frontières physiques pourrait avoir pour conséquence d'inciter les entreprises locales à employer des ressources distantes provenant d'autres pays, à moindres coûts.

Des interrogations importantes sont dès lors apportées par la tendance d'une adoption généralisée du télétravail : dans quelles mesures les employeurs devraient-ils couvrir les frais et les besoins des télétravailleurs ? Quelle serait la nouvelle compétitivité existante sur le marché globalisé de l'emploi et existerait-il un nouveau danger de délocalisation de l'emploi ? La motivation au télétravail serait-elle autant présente si les organisations venaient à mettre en place des outils de contrôle à distance de la performance plus rigoureux ? Plus globalement, est-ce que ces conséquences potentielles d'une adoption à long terme du télétravail seraient transposables aux conséquences de l'automatisation de procédés de travail, de l'adoption de technologies telle que l'intelligence artificielle, ou de la digitalisation des organisations en général ?

En dehors de l'attrait direct et court terme que peuvent procurer la technologie et la digitalisation des processus, la question des impacts sociaux et sociétaux sur des termes plus longs se pose. Comme le mentionnent Venkatesh et al. (2007), la recherche sur l'acceptation de la technologie devrait se focaliser sur d'autres questions importantes et mettre l'accent sur la dimension de la psychologie sociale. Les résultats de notre étude mettent finalement bien en avant l'importance moindre de l'aspect technologique dans l'adoption du télétravail face aux divers autres facteurs.

### 6.2     Limites et pistes de recherche

Plusieurs limites sont à relever pour cette étude. La première concerne l'échantillon de personnes interrogées, qui ne représente pas globalement la population de Suisse romande, mais uniquement celle concernée par le profil visé. La portée des résultats s'applique ainsi avant tout aux personnes ayant une activité professionnelle et dotées d'un niveau de formation équivalent ou supérieur à secondaire II.

Les résultats synthétisés sur la Figure 4 ne permettent pas de définir des liens de cause à effet, mais bien de mettre en lumière les facteurs explicatifs des variables de l'adoption du télétravail. En effet, le choix méthodologique d'utiliser la régression linéaire est limitant à ce niveau. De plus, la cohérence de l'homoscédasticité et de la normalité lors de l'utilisation des régressions linéaires a été vérifiée visuellement. Avec cette approche, nous ne sommes pas à l'abri d'un mauvais jugement, quand bien même une telle inspection visuelle est considérée comme tout à fait valide par les voix les plus autorisées (Salkind, 2016).

Enfin, citons encore la recherche des facteurs de télétravail, réalisée durant la phase qualitative de l'étude. Cette recherche s'est faite à partir de l'analyse d'un ensemble limité d'articles d'actualité grand public, ce qui ne permet pas d'exclure l'hypothèse de l'existence d'autres facteurs de motivation à télétravailler. Spécifiquement les problématiques de confiance (au sein notamment des collectifs de travail) et de sens du travail n'ont pratiquement pas été abordés ici, alors qu'elles pourraient également contribuer à expliquer l'adoption du télétravail (Hoornweg, Peters, & Van der Heijden, 2016; Visawanath Venkatesh & Speier, 2000). De même la prudence est de mise quand on sait que cette étude a été conduite en période de "télétravail forcé". Une situation susceptible de brouiller nos résultats en matière de motivation à adopter le télétravail, notamment en ce qui concerne la dimension intrinsèque de la motivation.

Bien que les résultats obtenus dans cette étude soutiennent une adoption du télétravail, une recherche approfondie, mettant en relation les modèles d'adoption (TAM) et les modèles de succès (ISSM), permettrait de présager sa pérennisation ou non dans le futur. D'autre part, il est aussi concevable que la Covid-19 ait créé des ponts et des raccourcis susceptibles d'avoir effacé certains obstacles, au moins temporairement. Dans ce cas, il serait intéressant de réitérer l'étude après la crise sanitaire, afin d'analyser dans quelles mesures ces obstacles auraient refait surface. Une telle étude permettrait, d'une part, de mieux objectiver les résultats actuels – les données auront alors été éprouvées en contexte de télétravail forcé et non forcé – et, d'autre part, de repérer les éventuelles différences entre catégories socioprofessionnelles. Ce qui permettrait d'éclairer davantage les raisons de l'adoption ou de la non-adoption du télétravail ; en particulier avec un nombre de répondants plus élevé.

# 7 CONCLUSION

Ce n'est pas un fait nouveau ; derrière la plupart des dispositifs techniques introduits dans les environnements professionnels se cachent des hommes et des femmes pour qui ces changements affectent le rapport au travail, respectivement la performance. Cette recherche revisite, dans une approche pragmatique, ce qui fait que le télétravail – dont on pourrait s'être, semble-t-il, trop vite réjoui – fonctionne concrètement. Au-delà de cette incontournable question, notre recherche pose aussi celle des conditions de succès lors du déploiement d'autres dispositifs techniques en lien avec la digitalisation des processus, l'adoption de l'intelligence artificielle ou l'automatisation de certains procédés de travail. Cette technicisation est appelée à gagner en importance dans la "nouvelle normalité" post-Covid-19 et se doit d'être débattue aujourd'hui. D'autre part, télétravailler efficacement rencontre plusieurs écueils susceptibles d'émousser la motivation. Au final, de quoi dépend cette dernière dans le contexte particulier du travail à distance. Notre réponse est double : d'abord l'attrait marqué pour le télétravail augmente dès lors que les employés en perçoivent une certaine utilité pour accomplir leur travail et ce d'autant plus si cela permet de gagner du temps de trajet domicile-travail. Ensuite l'adhésion et l'usage du télétravail restent intrinsèquement liés à la facilité d'utilisation des outils mis à la disposition des employés, ce qui implique un accompagnement spécifique dans le déploiement de cette nouvelle façon de travailler. À l'instar de la plupart des changements organisationnels, la réussite du télétravail nécessite de combiner sa dimension technologique aux défis liés au cœur social des entreprises. Un nouveau champ d'investissement et d'engagement pour les entreprises de demain ; l'utilité perçue et la facilité d'utilisation du télétravail n'allant pas de soi.

# 8 RÉFÉRENCES